\documentstyle[buckow]{article}

\def\cN{{\cal N}}

\def\be{\begin{equation}}
\def\ee{\end{equation}}

\newcommand{\bea}{\begin{eqnarray}}
\newcommand{\eea}{\end{eqnarray}}
\newcommand{\dalpha}{\dot{\alpha}}
\newcommand{\dbeta}{\dot{\beta}}
\newcommand{\dA}{\dot{A}}

\font\mybb=msbm10 at 12pt

\def\bb#1{\hbox{\mybb#1}}

\def\zet{{\bb{Z}}}

\def\real{{\bb{R}}}

\def\R4{\real^4}

\begin {document}
PAR--LPTHE-02/05
\def\titleline{D-instantons on orbifolds and gauge/gravity correspondence}
\def\authors{Alessandro Tanzini \1ad}
\def\addresses{
\1ad
LPTHE, Universit\'e Paris~VI, and INFN, Roma.\\
{\tt e-mail: tanzini@lpthe.jussieu.fr} \\}

\def\abstracttext{
D-instantons are used to probe the near-horizon geometry of
D3-branes systems on orbifold spaces. For fractional D3-branes,
D-instanton calculus 
correctly reproduces the gauge $\beta$-function and $U(1)_R$ anomaly of 
the corresponding ${\cal N}=2$ non-conformal Super Yang-Mills theories.
For D3-branes wrapping the orbifold singularity, D-instantons can 
be identified with gauge instantons on ALE space,
providing evidence of AdS/CFT duality for gauge theories on curved 
spaces.    }

\large
\makefront

The duality between ${\cal N}=4$ Super Yang--Mills theory (SYM) in four 
dimensions and type IIB string theory compactified on AdS$_5\times$S$^5$
\cite{mal} has provided a first concrete example of how a string theory
can be extracted from a gauge theory. A remarkable
result is that the multi--instanton moduli space of $SU(N)$ ${\cal N}=4$
SYM reproduces in the large $N$ limit the ten dimensional near-horizon
AdS$_5\times$S$^5$ geometry of the dual type IIB supergravity 
\cite{dhkmv}. This allows
to establish a careful correspondence between instanton dominated 
correlators in gauge theory and D-instanton corrections to the 
supergravity action, thus providing a non trivial check of the AdS/CFT 
duality. 

Recently, many interesting generalizations of this duality to 
more realistic non--confor-\break mal gauge theories have been 
considered.
In this context, the study of type IIB string theory on orbifold spaces 
play a major r$\hat{\rm o}$le. A general class of models with a lower 
amount of supersymmetry is in fact defined by locating a stack of 
D3-branes at a ${\real}^6/\Gamma$ singularity, where $\Gamma$ is a 
discrete subgroup of $SO(6)\cong SU(4)_R$. Depending on whether $\Gamma$
is embedded in an $SU(4), SU(3)$ or $SU(2)$ subgroup of $SU(4)_R$, the 
resulting low--energy Yang--Mills theories display respectively
$\cN=0,1,2$ supersymmetry \cite{kasi}. Moreover, if one consider 
fractional branes, which do not have images under the action of the group
$\Gamma$, the corresponding gauge theories turn out to be non-conformal 
\cite{kn}; various supergravity solutions of this kind have been studied 
for the ${\cal N}=2$ case \cite{dive}. However, in all these 
models the short distance region of the space-time is excised 
by an enhan\c{c}on mechanism, and additional states have to 
be taken into account to describe this region. 
In the gauge theory, the enhan\c{c}on
radius corresponds to the scale where instanton effects become 
relevant, thus one can expect that D-instantons come into play
on the supergravity side.
In \cite{fmt}, the low--energy dynamics of the D(-1)--D3 system in
presence of fractional branes have been considered,
along the lines of the analysis carried out in \cite{dhkmv,hkm}
for the conformal case.
The D--instanton partition function has been shown to factorize 
{\it already at finite $N$} in a term associated to the 
overall center of mass degrees of freedom, which describes the 
geometry of the space where the D--instantons are moving, 
and an effective matrix model describing internal degeneracies of the $k$ 
D--instantons bound state.  
In the present discussion, we focus on the pure $\cN=2$ SYM theory.
A more general analysis can be found in \cite{fmt}, 
to which we refer 
also for the notations and conventions and for a complete list of 
references.
The pure $\cN=2$ SYM action describes
the low--energy dynamics of a stack of $N$ fractional D3-branes lying at
a $\real^4/\Gamma$ singularity \cite{dive}.
This theory 
can be obtained by means of a $\Gamma$-projection on the 
${\cal N}=4$ $SU(N)$ gauge theory associated to $N$ nearby D3-branes 
moving on a flat space-time.
Analogously, the multi--instanton moduli space measure and 
multi--instanton action of pure $\cN=2$ 
can be read from the low--energy lagrangian describing the
dynamics of a fractional D(-1)--D3 system on $\real^4/\Gamma$.
We recall that the effective theory for $k$ D(-1) and 
$N$ D3-branes moving on flat space is given by the dimensional reduction
to $0+0$ dimensions of an ${\cal N}=1$ $U(k)$ gauge theory
in $D=6$ with vector multiplet $(\chi_a,\lambda_{\dalpha A},D_c)$,
one hypermultiplet 
transforming in the adjoint 
representation $(a_{\alpha\dalpha},{\cal M}_\alpha^A)$
and $N$  transforming in the fundamental representation (and its 
conjugate),
denoted by $(w_{\dalpha},\mu^A ;\bar{w}^{\dalpha},\bar{\mu}^A)$
\footnote{Here $a=1,\ldots,6$ stands for the vector index in the $SO(6)$
group acting on the transverse space to the D3-branes, 
$\alpha,\dalpha=1,2$ for spinorial indices of the Lorentz group acting
on the longitudinal space, and $A=1,\ldots,4$ for the 
fundamental index of the internal $SU(4)_R\cong SO(6)$ group. Finally, 
$c=1,2,3$ labels 
the auxiliary fields of the vector multiplet \cite{fmt}.}.
The low--energy excitations of the fractional D(-1)--D3 system are defined
by the $\Gamma$--projection on these fields, which for 
$\Gamma=\zet_p$ reads\footnote{We adopt the notations of \cite{hkm}.
Given the local group isomorphism $SO(6)\cong
SU(4)$, $SO(6)$ vectors can also be written as
$\chi_{AB}\equiv \Sigma^a_{AB} \chi_a / \sqrt{8}$, 
with $\Sigma^a_{AB}$ given in terms of the t'Hooft symbols.}
\bea
\mu^A &=& e^{2\pi i {q_A \over p}} \mu^A \quad\quad
\bar{\mu}^A = e^{2\pi i {q_A \over p}}
\bar{\mu}^A \nonumber\\
{\cal M}^A_{\alpha} &=& e^{2\pi i {q_A \over p}}
{\cal M}^A_{\alpha} \quad\quad
{\lambda}^A_{\dalpha} = e^{2\pi i {q_A \over p}}
{\lambda}^A_{\dalpha} \nonumber\\
\chi^{AB} &=& e^{2\pi i {q_A+q_B \over p}} \chi^{AB}   
\label{Ginv}
\eea
with $q_1=q_2=0$, $q_3=-q_4=1$. All the other fields
$(w_{\dalpha},\bar{w}^{\dalpha},a_m,D^c)$
are invariant under the $\zet_p$ orbifold action.
The net effect of the projection (\ref{Ginv}) on the bosonic fields is 
to set to zero the components $(\chi_1,\chi_2,\chi_4,\chi_5)$ of the 
vector field $\chi_a$, which roughly describes the position of the 
D-instantons in the transverse space to the D3-branes. Thus the 
fractional D-instantons are allowed to move only on the orbifold plane
$(\chi_3,\chi_6)$.
As for the fermionic fields, it is easily realized 
that the projection (\ref{Ginv}) breaks the $SU(4)$
${\cal R}$-symmetry group of the ${\cal N}=4$ theory down to
$SU(2)_{\dot{A}}\times U(1)_{\cal R}$,
where $SU(2)_{\dot A}$ is the automorphism
group and $U(1)_{\cal R}$ the anomalous ${\cal R}$ charge 
of the $\cN=2$ SYM theory.
By introducing the complex coordinates on the orbifold plane
\bea
\phi &\equiv& 2\chi_{43}=2\chi^{12}={1\over \sqrt{2}} (-\chi_3+i \chi_6)
\nonumber\\
\bar{\phi} &\equiv& 2\chi_{21}=2\chi^{34}={1\over \sqrt{2}} (-\chi_3-i 
\chi_6) \ \ ,
\label{phi}
\eea
we can now write the $\cN=2$ multi--instanton action 
as
\be
S_{k,N}={1\over g_0^{2}}S_{G} + S_{K}+S_{D}
\label{cometipare}
\ee
with
\bea
&&S_{G}={\rm tr}_{k}\Big(2[\phi,\bar\phi]^2+{i\pi\over\sqrt{2}}
\lambda_{\dot{\alpha} 
\dot{A}}[\phi,\lambda^{\dot{\alpha}\dot{A}}]
-D^{c}D^{c}\Big)
\label{Sd}\\
&&S_{K}={\rm tr}_{k}\Big(2 \bar\phi L \phi 
- \sqrt{2}i\pi \bar\phi (\bar\mu_{\dot{A}}\mu^{\dot{A}}
+ {\cal M}^\alpha_{\dot{A}}{\cal M}_\alpha^{\dot{A}}) \Big)
\, \nonumber\\
&&S_{D}={\rm tr}_k\,\Big(i \pi (
-[a_{\alpha\dot{\alpha}},{\cal M}^{\alpha \dot{A}}]
+\bar{\mu}^{\dot{A}} w_{\dot{\alpha}}
+\bar{w}_{\dot{\alpha}}\mu^{\dot{A}})
\lambda^{\dot{\alpha}}_{\dot{A}}
+D^{c}\left(\bar{w} \tau^c w-i \bar{\eta}_{mn}^c [a_{m},a_{n}]
\right)\Big)\nonumber \ \ ,
\eea
where we have defined the linear operator 
$L\cdot \Omega = \{W^0,\Omega\}/2 + [a_m,[a_m,\Omega]]$,
with $W^0_{ij}=\bar{w}^{\dalpha}_{i u}w_{u j \dalpha}$.
It is worth to remark that the multi--instanton action (\ref{cometipare}) 
can be easily twisted 
into a topological matrix model \cite{fmt}. Thus, the parameters 
appearing 
in this action can be continuously deformed without affecting
the results on the computation of the observables, as long as they are 
well defined. In particular, the partition function can be evaluated in 
the limit $g_0\rightarrow\infty$; life is much 
simpler in this limit, since the term $S_G$ in (\ref{cometipare}) is 
suppressed and the fields $(D_c,\lambda_{\dalpha}^{\dot{A}})$ become
lagrangian multipliers implementing the ADHM constraints in 
$S_D$ \cite{dhkmv}.
In order to explicitly evaluate the partition function it is moreover 
convenient to define the 
$SU(N)$ gauge invariant variables $W^m_{ij},\zeta^{\dalpha \dot{A}}_{ij}$
through \cite{dhkmv}
\bea
W^0_{ij} &=&\bar{w}_{iu}^{\dalpha}
 \,w_{uj\dalpha}\ ,\quad \quad\quad\quad\quad 
W^c_{ij}=\bar{w}_{iu}^{\dalpha}   
\, \tau^c {}^{\dbeta}_{\dalpha} \,w_{uj\dbeta}~~~ c=1,2,3\nonumber\\
\mu_{iu}^{\dot{A}} &=& w_{uj\dalpha}\zeta^{\dalpha \dot{A}}_{ji}+
\nu_{iu}^{\dot{A}},\quad\quad\quad
{\rm with}~~\bar{w}_{iu}^{\dalpha} \nu_{uj}^{\dot{A}}=0, \nonumber\\
\bar{\mu}_{iu}^{\dot{A}} &=& \bar{\zeta}^{\dot{A}}_{\dalpha ij}
\bar{w}_{uj}^{\dalpha}
+\bar{\nu}_{iu}^{\dot{A}}\quad\quad\quad\quad
 {\rm with}~~\bar{\nu}_{iu}^{\dot{A}}
 w^{}_{uj\dalpha}=0\
\label{invv}
\eea
and perform the integrations over the iso-orientation
modes (parameterizing a point in the
coset space ${SU(N) \over SU(N-2k)\times U(1)}$) and over their
``fermionic superpartners'' $\nu^{\dot{A}}$, $\bar{\nu}^{\dot{A}}$.
Then by integrating out the auxiliary
field $D^c$ we enforce the bosonic ADHM constraint 
$W^c=i\bar{\eta}^c_{mn}[a_n,a_m]$,
while the integration on $\lambda_{\dalpha}^{\dA}$ enforces the fermionic
constraint. After these manipulations, we are finally left with
the $SU(N)$ invariant measure \cite{hkm}
\bea
Z_{k,N} &=& c_{k,N}
e^{2\pi i k \tau}
{1 \over {\rm Vol}\, U(k)} \int d^{k^2} W^0\, d^{4k^2}a \,
d^{k^2}\phi\,d^{k^2}\bar{\phi}\,
d^{4k^2}{\cal M} \, d^{4k^2}\zeta \,
\nonumber\\
&& \times \left({{\rm det}_{2k} W} {\rm 
det}_{k}\bar{\phi}^2\right)^{N-2k}\,
e^{-S_{k,N}} \ \ ,
\label{zkn2}
\eea
with
\bea
S_{k,N} &=& - 4\pi i \,{\rm tr}_k\,\bar{\phi}
{\Lambda}^{12}+ {\rm tr}_k\,(2 \bar{\phi}L \phi) \, 
\label{l2}\\
{\Lambda}^{\dot{A}\dot{B}}&=&{1\over{2\sqrt{2}}}
\left(\zeta^{[\dot{A}} W \zeta^{\dot{B}]}
+[a^{\alpha \dalpha}, {\cal 
M}_{\alpha}^{[\dot{A}}]\zeta_{\dalpha}^{\dot{B}]}
+{\cal M}^{\alpha [\dot{A}}{\cal M}_{\alpha}^{\dot{B}]}\right) 
\nonumber \\
c_{k,N} &=&
{2^{2kN-4k^2+k}\pi^{2kN-2k^2+k}\over\prod_{i=1}^{2k}(N-i)!}
(-2\pi^2)^{k(N-2k)}\pi^{4k^2}
 \ \ .
\nonumber
\eea
The determinant factors in (\ref{zkn2}) come from the Jacobian of the
change of variables (\ref{invv}) and from the integration on the fermionic
variables $\nu^{\dot{A}}$, $\bar{\nu}^{\dot{A}}$.

We would like now to factor out the dependence on the center of mass
degrees of freedom in (\ref{zkn2}) in order to get informations on the
near--horizon geometry of our model. As we will see shortly, this can be
done already at finite $N$. Inspired by large $N$ manipulations of 
\cite{dhkmv} we
split each $U(k)$ adjoint V-field into its trace and traceless part
\be
V= v_0 {\bf 1}_{k\times k} + \hat{v} \ \ .
\ee
At this point it is convenient to rescale the various
traceless components of the fields in the following
way
\bea
a_m &=& -x_m+ \rho \hat{a}_m
\nonumber\\
\phi &=& r e^{2i \vartheta}\left(1+\hat{\phi}\right)
\nonumber\\
\bar{\phi} &=& r e^{-2i \vartheta}\left(1+\hat{\bar{\phi}}\right)
\nonumber\\
W_0 &=& \rho^2 \left(1+\hat{W}_0\right)
\nonumber\\
\zeta_{\dalpha}^{\dot{A}} &=& e^{i\vartheta} 
\bar{\eta}_{\dalpha}^{\dot{A}}
+\rho^{-{1\over 2}} e^{i\vartheta}
\hat{\zeta}_{\dalpha}^{\dot{A}}\nonumber\\
{\cal M}_{\alpha}^{\dot{A}} &=& e^{i\vartheta} 
{\xi}_{\alpha}^{\dot{A}}
+ \rho \hat{a}_{\alpha \dalpha}e^{i\vartheta} 
\bar{\eta}^{\dalpha \dot{A}} +   
\rho^{{1\over 2}} e^{i\vartheta}\hat{\cal{M}}_{\alpha}^{\dot{A}} \ \ .
\label{splitting}
\eea
From (\ref{splitting}) it follows that the variables $x_m$, $\rho^2$,
${\xi}_{\dalpha}^{\dA}$, $\bar{\eta}_{\dalpha}^{\dot{A}}$ are
the trace components associated
to the overall bosonic and fermionic translational/conformal zero modes,
while $r$ is the distance of the D-instanton probe from the D3-branes
\footnote{Actually $r=d/\alpha^{\prime}$, where $d$ is the distance from
the D3-branes in the orbifold fixed plane.
}.
Our results can be expressed in terms of this quantity and of
the adimensional variable $y=r\rho$.
In fact, plugging (\ref{splitting}) in (\ref{zkn2}) and taking care
of the Jacobian involved in the rescalings, one can easily check that
all dependence in the center of mass variables
$r, x_m, \vartheta$ factors out as 
\bea
Z_{k,N} &=& d_{k,N} \int
d^{4}x_m \, r^3 dr d\vartheta \,
e^{-4 i k N\vartheta} r^{-2kN}
e^{2\pi i k \tau_0} \,
d^{4}{\xi} \, d^{4}\bar{\eta}
\nonumber\\
&=&d_{k,N} \int
d^{4}x_m \, r^3 dr d\vartheta \,
e^{2\pi i k \tau(r)} \,
d^{4}{\xi} \, d^{4}\bar{\eta} \ \ .
\label{partfuncn=2}
\eea
In (\ref{partfuncn=2}) we can easily recognize the 
AdS$_5\times$ S$^1$ volume form
\footnote{We use coordinates \cite{mal}
where the AdS$_5\times$ S$^1$ metric
reads $ds^2={r^2\over R^2} d x_m^2+{R^2 \over r^2} dr^2 +R^2 d\vartheta^2$
with $R^2=\sqrt{4\pi g_s N}$, $g_s$ being the string coupling constant.},
$d^4 x_m \, r^3dr d\vartheta$,
multiplied by a power of the complex coordinate $re^{2i\vartheta}$. 
Remarkably, this deformation of the AdS$_5\times$S$^1$
factor can be completely reabsorbed 
in the following redefinition of the complex coupling
\be
2\pi i \tau(r,\vartheta)\equiv  i {\rm \theta} - {8\pi^2 \over g(r)^2}
= i ({\rm\theta}_0 - 4N\vartheta) - {8\pi^2\over g_0^2}- 2N \, {\rm 
ln} (r)
\ \ .
\label{pi}
\ee
If, as usual in AdS/CFT correspondence, we identify
the AdS radial coordinate $r$ with the dynamical energy scale $\mu$
of the SYM gauge theory, we can read from the term multiplying the 
logarithm in (\ref{pi})
the first coefficient in the expansion of the $\beta$ function
for the gauge coupling
\be
\beta\equiv r {d \over dr} g(r)=- b_1 g^3/16\pi^2+\ldots \ \ .
\label{beta}
\ee
with $b_1=2N$, in agreement with the field theory result.
Moreover,
the extra phase factor $ e^{-4 i k N  \vartheta}$ in 
(\ref{partfuncn=2}) correctly reproduces the $U(1)_R$ anomaly of
$\cN=2$ SYM. Notice in fact that the angular rescalings in
(\ref{splitting}) correspond to an $U(1)_R$ transformation
on the fields of the $\cN=2$ SYM.
Due to the presence of instantons, such a transformation is not
a symmetry of the lagrangian, but induces a variation of the 
${\rm \theta}_0$ parameter precisely
equal to that we have found in (\ref{pi}). This also imply 
that in order to have a non--trivial correlation function one should 
consider the insertion of suitable operators carrying $n=4kN$ fermionic
zero--modes. An interesting example could be the   
single or multi-trace composite operators of scalar fields, 
which have been shown not to get perturbative anomalous dimensions in
${\cal N}=2$ theories \cite{Maggiore:2001}.
Finally, the $d_{k,N}$ in (\ref{partfuncn=2}) are numerical coefficients 
given by integrals on the left--over $SU(k)$ variables, which describe the
internal dynamics of the $k$ D--instanton bound state. 
They can be computed only in a limited number of cases, and notably
in the large $N$ limit where they are an important ingredient to check the
AdS/CFT duality \cite{dhkmv}. The formulation in terms of topological
matrix theory is a crucial tool in this evaluation, and could be useful 
also for finite $N$ computations. However, we remark that the coefficients
$d_{k,N}$ do not involve powers of $r$, and therefore cannot affect the
result found in (\ref{partfuncn=2}).

Summarizing, the near--horizon geometry turns out to be a deformed 
AdS$_5\times$S$^1$, where the deformation is entirely determined in 
terms of the one--loop $\beta$ function and $U(1)_R$ anomaly of the 
underlying gauge theory.
By means of the technique just described one can study  
also more general configurations containing regular D3-branes 
and regular D-instantons. Since these last 
possess images under the discrete group $\Gamma$, they can
leave the orbifold plane and roam in the entire transverse 
space, probing the complete AdS$_5\times$S$^5/\zet_p$
geometry \cite{fmt}.


Let us now comment on the case in 
which the orbifold quotient is taken along the directions longitudinal to the
D3-branes system \cite{dmjm}. This means that the discrete group $\Gamma$ 
is now embedded in an $SU(2)$ factor of the Lorentz group $SO(4)\cong 
SU(2)_L \times SU(2)_R$, say for definiteness $SU(2)_L$. The projection is thus 
similar to that in (\ref{Ginv}), but now $\Gamma$ is acting on the spinor
indices $\alpha,\beta$ of $SU(2)_L$. Moreover, the orbifold group acts
non trivially also on the Chan--Paton indices. Thus, starting
from a solution with instanton number $K=\sum_q k_q$ and gauge group
$U(N)$, $N=\sum_q N_q$, the surviving components 
turn out to be \cite{fmt}
\bea
w_{\dalpha}^q &=& w_{\dalpha i_q u_q}\quad\quad
\mu^A_{q} = \mu^A_{i_{q} u_q} \nonumber\\
a^{\alpha\dalpha}_q &=& a^{\alpha\dalpha}_{i_{q+q_\alpha} j_q}\quad\quad
{\cal M}^{A\alpha}_q = {\cal M}^{A\alpha}_{i_{q+q_\alpha} j_q} \nonumber\\
\chi^{AB}_q &=& \chi^{AB}_{i_{q} j_{q}} \quad
D^c_q= D^c_{i_q j_q}
\quad
\lambda^{A\dalpha}_q = \lambda^{A\dalpha}_{i_{q} j_q} \ \ .   
\label{invale}
\eea
They correspond to block matrices with block indices 
$i_q=1, \ldots,k_q$, $u_q=1,\ldots,N_q$, and
$q=0,\ldots,p-1$ for $\Gamma=\zet_p$.
All sums in 
$q$ are from now on understood modulo $p$.
Notice that, contrary to the
previous case (\ref{Ginv}), the whole $\cN=4$
supersymmetry is preserved by this projection since $\Gamma$ acts in 
the same way on the different components of a given supermultiplet.
The moduli space of multi-instanton solutions
and ADHM constraints can be described 
in terms of the invariant components (\ref{invale}).
In particular, the moduli space is spanned by the super--coordinates
in the first two lines of (\ref{invale}). The presence of 
${\rm dim}\, D^c=3 k_q^2$ bosonic ADHM constraints for each $q$ and of
the auxiliary gauge group $\prod_q U(k_q)$ reduces the number of
independent bosonic degrees of freedom to
\be
{\rm dim}\, {\cal M}_B = 4\sum_q (k_q N_q+\hat{k}_q k_q-k_q^2) \ \ ,
\label{mb}
\ee
where $\hat{k}_q={1\over 2} \sum_{\alpha} k_{q+q_\alpha}$,
with $q_1=-q_2=1$
Analogously, due to the presence of 
${\rm dim}\, \lambda^{A\dalpha}=8k_q^2$ fermionic ADHM constraints for 
each $q$, 
the number of independent fermionic degrees of freedom is 
\be
{\rm dim}\, {\cal M}_F = 8\sum_q (k_q N_q+\hat{k}_q k_q-k_q^2) \ \ .
\label{dimension}
\ee
The resulting dimensions of the moduli space in 
(\ref{mb}) and (\ref{dimension}) are in agreement
with that obtained from the Kronheimer--Nakajima 
construction of gauge instantons on ALE spaces \cite{Bianchi}.
Thus, in presence of D3-branes wrapping an $\real^4/\zet_p$ singularity, 
we are led, following \cite{dmjm}, 
to identify D-instantons with self-dual connections for non--abelian gauge 
theories living on ALE spaces.
An important new feature emerges in this case;
indeed, since ALE manifolds support non trivial 
two--cycles, the first Chern class is in general non--vanishing and 
characterizes the instanton solution together with the second Chern
class usually considered on flat spaces. 
In particular, instanton solutions with vanishing first 
Chern class must obey the condition \cite{fmt}
\be
N_q+2k_q-2\hat{k}_q=0 \quad\quad {\rm for}~~~q>0.
\label{c10}
\ee
We remark that only in these cases the second Chern class 
coincides with the instanton number defined as $K/|\Gamma|=K/p$.
(\ref{c10}) is a highly non trivial constraint on the allowed values of 
$(k_q,N_q)$. In the simplest case, {\it i.e.} the
$SU(2)$ bundle on the Eguchi--Hanson blown--down space $\real^4/\zet_2$,
solutions to (\ref{c10}) are given by either $\vec{N}=(2,0), 
\vec{k}=(k,k)$ or
$\vec{N}=(0,2), \vec{k}=(k-1,k)$. The corresponding instanton solutions
have respectively integer and half--integer second Chern class, while the 
dimension of the moduli space computed from (\ref{mb})
turns out to be equal to $8k$ and $8k-4$ respectively, in agreement with
\cite{Bianchi}.
A computation of the partition function for the lowest value of the
Chern class, $c_2=1/2$, was carried out in \cite{Bianchi}
yielding the bulk contribution to the Euler number of the
moduli space\footnote{It is known and rigorously proven that for
$c_2=1/2$ the moduli space is a copy of the base manifold (the   
Eguchi-Hanson manifold) whose bulk contribution is $3/2$.}.

The low energy action for this D(-1)--D3 system
can be easily obtained by implementing the $\Gamma$-projection
(\ref{invale}) on the $U(k)$ effective gauge theory of the parent 
D(-1)--D3 system living on flat space.
Since the whole $SO(6)$ ${\cal R}$-symmetry is clearly
preserved by the projection, this action describes
instantons in an ${\cal N}=4$ gauge theory,
living on an ALE space.
It is a straightforward exercise to
apply the techniques previously described to show that
for a regular D-instanton probe the center of mass degrees of freedom factor 
out from the multi-instanton measure and describe a point in an
euclidean $S^5\times AdS^E_5/\zet_p$ space, with $\zet_p$ acting on the 
four-dimensional $\real^4$ boundary of the $AdS^E_5$ \cite{fmt}.


\end{document}